\documentstyle[emulateapj]{article}

\begin{document} 
 
\title{Gravitational Evolution of the Large-Scale Probability Density
Distribution: The Edgeworth \& Gamma Expansions}

\author{E. Gazta\~{n}aga $^{1}$, P. Fosalba $^{1,2}$, E. Elizalde $^{1}$}

\affil{${}^1$ Consejo Superior de Investigaciones Cient\'{\i}ficas (CSIC), 
Institut d'Estudis Espacials de Catalunya (IEEC), \\ 
Edifici Nexus-201, Gran Capit\`a 2-4, 08034 Barcelona, Spain \\}
\affil{${}^2$ Astrophysics Division, Space Science Department 
of ESA, ESTEC, NL-2200 AG Noordwijk, The Netherlands}

%\authoremail{gaztanaga@ieec.fcr.es, pfosalba@astro.estec.esa.nl, elizalde@ieec.fcr.es} 
 
%\maketitle 
 
\def\Mpc{{\,h^{-1}\,{\rm Mpc}}} 
\def\mpc {h^{-1} {\rm{Mpc}}} 
\def\and  {\it {et al.} \rm} 
\def\rmd {\rm d} 
 
\def\la{\mathrel{\mathpalette\fun <}}
\def\ga{\mathrel{\mathpalette\fun >}}
\def\fun#1#2{\lower3.6pt\vbox{\baselineskip0pt\lineskip.9pt
\ialign{$\mathsurround=0pt#1\hfill##\hfil$\crcr#2\crcr\sim\crcr}}}

\def\etal{{\rm et~al. }} 
\def\spose#1{\hbox to 0pt{#1\hss}} 
\def\simlt{\mathrel{\spose{\lower 3pt\hbox{$\mathchar"218$}} 
     \raise 2.0pt\hbox{$\mathchar"13C$}}} 
\def\simgt{\mathrel{\spose{\lower 3pt\hbox{$\mathchar"218$}} 
     \raise 2.0pt\hbox{$\mathchar"13E$}}} 
\def\beq{\begin{equation}} 
\def\eeq{\end{equation}} 
\def\bce{\begin{center}} 
\def\ece{\end{center}} 
\def\bea{\begin{eqnarray}} 
\def\eea{\end{eqnarray}} 
\def\ben{\begin{enumerate}} 
\def\een{\end{enumerate}} 
\def\ul{\underline} 
\def\ni{\noindent} 
\def\nn{\nonumber} 
\def\bs{\bigskip} 
\def\ms{\medskip} 
\def\wt{\widetilde} 
\def\brr{\begin{array}} 
\def\err{\end{array}} 
\def\dsp{\displaystyle} 
\newcommand{\rhobar}{\overline{\rho}} 
\newcommand{\xibar}{\overline{\xi}} 
\def\Or{{\cal O}} 
%%%%%%my def: EG%%%%%%%%%%%%%%%%%%%%%%%%%%%%%%%%%%%%%%%%%%%%%%%%%%%%%%%%%% 

\font\twelveBF=cmmib10 scaled 1200 
\newcommand{\bte}{\hbox{\twelveBF $\theta$}} 
\newcommand{\x}{\hbox{\twelveBF x}} 
\newcommand{\X}{\hbox{\twelveBF X}} 
\newcommand{\vv}{\hbox{\twelveBF v}} 
\newcommand{\y}{\hbox{\twelveBF y}} 
\newcommand{\r}{\hbox{\twelveBF r}} 
\newcommand{\k}{k} 
\newcommand{\lexp}{\mathop{\bigl\langle}} 
\newcommand{\rexp}{\mathop{\bigr\rangle}} 
\newcommand{\rexpc}{\mathop{\bigr\rangle_c}{}} 
\newcommand{\eq}{{equation~}} 
\def\v{\vec{v}}	% velocity vector 
\def\del{\delta}   	% cosmic field 
\def\N{N}		% N-points 
\def\M{M}		% M-points 
\def\J{J}		% J-order of moment 
\def\I{I}		% I-order of moment 
\def\calE{\cal{E}}      % event 
\def\Omeg{{\cal{S}}}	 % sample space 
\def\F{F}		% cumulative distribution 
\def\V{{\cal{V_T}}}	% volume integral 
\def\m{m}		% 1-point moments 
\def\muc{{\mu}}		% 1-point central moments 
\def\kum{k}		% cumulants 
\def\barN{\bar{N}}		% mean counts-in-cells 
\def\calD{\cal{D}}		% functional measure 
\def\Mgf{{\cal{M}}}		%  moment generating function 
\def\Mgff{{{M}}}		%  functional moment generating function 
\def\psic{{\cal{\psi}}}		%  cumulant generating function 
\def\psif{{\psi}}		%  functional generating function 
\def\B{{\cal{B}}}		% dimensional ratios of cumulants 
\def\P{{\cal{\Xi}}}		% multipont in fourier 

\begin{abstract} 
 
The gravitational evolution of the cosmic 
one-point probability distribution function 
(PDF)  has been estimated using an analytic approximation  
that combines gravitational perturbation theory 
with the Edgeworth expansion around a Gaussian PDF.
Despite the remarkable success of the 
Edgeworth expansion in modeling the weakly non-linear growth of
fluctuations around the peak of the cosmic PDF, it fails to reproduce
the expected behaviour in the tails of the distribution.  
Besides, this expansion is ill-defined as it predicts negative densities
and negative probabilities for the cosmic fields.
This is a natural consequence of using an expansion around the Gaussian
distribution, which is not rigorously well-defined 
when describing a positive variate, 
such as the density field. 
Here we present an alternative to the Edgeworth  
series based on an expansion around the Gamma PDF. 
The Gamma expansion is designed to converge when the 
PDF exhibits exponential tails, which
are predicted by Perturbation Theory, in the weakly non-linear regime,
and are found in numerical simulations from Gaussian initial conditions.
The proposed expansion is better suited for describing a real PDF 
as it always yields positive  
densities and the PDF is effectively positive-definite. 
We compare the performance of the Edgeworth and the Gamma
expansions for a wide dynamical range making use 
of cosmological N-body simulations and assess their range of validity.
In general, the Gamma expansion provides an interesting and simple
alternative to the Edgeworth series and it should be useful 
for modeling non-gaussian PDFs in other contexts, such as in the
cosmic microwave background.
 
\end{abstract} 
 
%\keywords{cosmic microwave background --- cosmology:observations --- 
%methods:statistical} 
 
\section{Introduction} 
 
We aim at studying one dimensional probability density 
functions (PDF), $p(\del)$, which characterize the 
statistical properties of a stochastic field at a single point 
$\del=\del(\r)$. 
Here we shall concentrate on fluctuations of the  
density field: 
$\del \equiv \rho/\bar{\rho} -1$ (being $\bar{\rho}$ the mean value of the 
density field, $\rho$), smoothed over some fixed 
scale $R$. However many of the arguments presented here are 
quite generic and could also be applied to other contexts. 
The approach we will follow  is to try to recover the  
full shape of the PDF from the knowledge of its first order  
moments.  
This has become by now a textbook problem and there are several  
standard ways to address it (for a review see 
Kendall, Stuart \& Ord 1994). The solution  is not unique, 
unless there is a well defined hierarchy in the moments and 
we can define some perturbative approach to the problem. 
 
There are a number of studies to predict the evolution of clustering 
of density fluctuations, and in particular of its PDF. 
There have been attempts to derive the PDF from analytic approximations,
such as the Zeldovich Approximation 
(Kofman et al. 1994). 
Although the Zeldovich Approximation reproduces important aspects of the non-linear 
dynamics, it is a poor approximation for the PDF and its moments. 
One way to improve that is to take advantage of the exact non-linear 
perturbation  theory (PT) to estimate the moments (Bernardeau 1992) and use them 
to derive the PDF from the Edgeworth 
expansion (Juszkiewicz et al 1995, Bernardeau \& Koffman 1995). 
In this case the PDF is predicted to an  
accuracy given by the order of the cumulants involved. 
The Edgeworth expansion has since been used as a tool to characterize
the PDF of matter (eg Kim \& Strauss 1998, Blinnikov \& Moessnew 1998, Pen 1998) and
CMB  fluctuations (Amendola 1996, Popa 1998).
Earlier phenomenological approaches to the construction of the 
cosmic PDF were developed by
Saslaw \& Hamilton (1984) and Coles \& Jones (1991).

One serious shortcoming of the Edgeworth
approach is that the series yields 
a PDF that is ill-defined. It has negative probability 
values and assignes non-zero 
probability to negative densities ($\del<-1$). This latter problem  
originates on the fact that the Gaussian PDF, which is the basis for the 
Edgeworth series, only makes physical sense when the rms fluctuation 
$\sigma$ is very small. Here we will try  
to address some of these problems by exploring the possibility of carrying out 
expansions around better behaved PDFs, 
more suitable to yield positive densities 
when the variance is not that small. We shall concentrate on the 
Gamma PDF, but other distributions may be handled within the same framework, 
as our general analysis will show.

Whenever the moment generating function of the PDF is known, 
it is then possible  to reconstruct the full PDF. 
This has been done previously by using the Legendre 
transform (Fry 1985) or the inverse Laplace transform 
(Balian \& Schaeffer 1989, Bernardeau 1992, Bernardeau \& Kofman 1995). 
Since for gravitational clustering in the weakly non-linear regime
the variance of the PDF is small, one can
expand the above transforms to recover the PT limit.
In the case of Gaussian initial fluctuations this perturbative
 expansion yields to the well-known Edgeworth series.

In \S\ref{sec:expand} we shall introduce the Edgeworth 
series as a generic saddle point 
approximation to the PDF, along the lines of Fry (1985).
An alternative expansion, around the Gamma distribution, 
is introduced in  \S\ref{sec:gamma}. It is expressed as 
an orthogonal polynomial expansion of an arbitrary PDF
in terms of Laguerre polinomials. The latter are the 
counterparts to the Hermite 
polinomials when one expands around an exponential 
tail instead of a Gaussian one.
A detailed comparative analysis of the Edgeworth 
and Gamma expansions with respect to 
N-body simulations is presented in 
\S\ref{sec:compare} and \S\ref{sec:nbody}.
A final discussion with our conclusions is given in  
\S\ref{sec:conc}.

\section{Expansions around a given PDF} 
\label{sec:expand} 
 
\subsection{One-Point Statistics}

As usual, we shall denote statistical averaging  
by brackets: $\lexp ... \rexp$, so that the expectation value 
for the moments are: 
 
\beq 
m_\J \equiv \lexp \del^\J \rexp = \int p(\del) \del^\J d\del 
\eeq 
with $\J$ an integer that labels the order of the corresponding 
moment. $\J=1$ corresponds to the mean, which for the density 
fluctuation is zero, $m_1=\lexp \del \rexp=0$.  
In this notation the variance, Var $(\del)$, and 
{\it rms} fluctuation $\sigma$, are defined as: 
Var $(\del) \equiv \sigma^2 \equiv m_2 - m_1^2$. 
It is useful to introduce the {\it cumulants} $\k_\J$: 
\beq 
\k_J\, \equiv \lexp \del^J \rexpc  
= ~ \left.{\frac{d^{J} \psic(t) }{dt^{J}}}\right|_{t=0} 
= ~ \left.{\frac{d^{J} \log{\Mgf(t)} }{dt^{J}}}\right|_{t=0}, 
\label{cgf} 
\eeq 
where $\psic(t)\equiv  \log{\Mgf(t)}$  
is given in terms of the moments of the PDF through  
$\Mgf(t)$: 
 
\beq 
\Mgf(t) \equiv \lexp e^{t\del} \rexp ~=~ \int_{-\infty}^{\infty}  p(\del)\,e^{t\del}\,d\del = \sum_{\J}  \, {t^\J \over{J!}} \, m_\J 
\eeq

Gravitational clustering from Gaussian initial 
conditions predicts ${ <\del^J >_{c}} \propto { <\del^2 >_{c}}^{\J-1}$
on large-scales, thus it is more convenient to introduce the following ratios,    
\beq 
S_{\J}\, \equiv \,\frac{\k_\J}{{\k_2^{\J-1}}} = { \lexp \del^J \rexpc\over{ \lexp \del^2 \rexpc}^{\J-1}}. 
\label{sj} 
\eeq 
where the {\it Skewness}, $S_3$, is the third-order ratio, and the {\it Kurtosis}, $S_4$,
is the fourth-order one. 
From $\psic(t)$ we can get back the PDF, $p(\del)$, by 
using the inversion formula: 
\beq 
p(\del) =  \int_{-i \infty}^{ +i \infty}  
{dt \over {2 \pi}}e^{t\del+\psic(t)}.  
%= 
%\int_{-i \infty}^{ +i \infty}  
%{dt \over {2 \pi}}~\exp{\left\{t\del+\sum_{\J}  \, {t^\J \over{J!}} \,  
%\k_\J \right\} }. 
\label{PDF-mgf} 
\eeq 
 
  Consider two differentiable 
PDFs, $p_1(\del)$ and $p_2(\delta)$, with cumulants 
$\kum_\J^{(1)}$ and $\kum_\J^{(2)}$,  
it follows that (see eg. Kendall, Stuart \& Ord 1994): 
\beq 
p^{(1)}(\del) = \exp\left[ \sum_{J=0}^\infty ~(-1)^J~{\kum_J^{(1)}-\kum_J^{(2)}\over{J!}} 
~ {d^J\over{d\del^J}}  \right] ~ p^{(2)}(\del). 
\label{p1p2} 
\eeq 
This equation is easy to prove by reobtaining the moments of $p^{(1)}$ 
 through the generating 
function, having assumed that those of $p^{(2)}$ are given by $k_J^{(2)}$. After partial  
integration, the generating function of $p^{(1)}$ arises immediately, what proves  
the above equality (it is a nice exercise). 
 
Equation (\ref{p1p2}) allows one to use the most 
convenient PDF to do the series expansion, $p^{(2)}(\delta )$. 
In particular, if one uses the Gaussian as the parent PDF one ends up with the 
Gram-Charlier series which yields the Edgeworth expansion in the perturbative limit 
(ie, when the variance of the PDF is small).

In this paper, we will focus on the Gamma distribution.
In that case, we have that
$p^{(2)}(\delta )$ is given by the Gamma PDF (see Eq(\ref{gamPDF}) below)
with $\delta =\beta z + \alpha$. 
Naming $p^{(i)}(\delta )=\bar{p^{(i)}} (z)$ ($i=1,2$), we get 
\beq 
{\bar{p^{(1)}}(z)\over{\bar{p^{(2)}} (z)}}
 = \exp \left[ \sum_{J=0}^\infty  (k_J^{(1)}-k_J^{(2)})\, 
\left( \frac{-1}{\beta z} 
\right)^J L_J^{(p-J-1)} (z)\right]
\label{p1p2b}
\eeq 
However such a straightforward approach is ill-defined as the measure of the
expansion is proportional to the order considered 
(the order of the expansion, $J$, 
is involved in the index denoting the order 
of the generalised Laguerre polinomials
in Eq.[\ref{p1p2b}]).
 
This will force us to modify the general method to derive
a consistent expansion in terms of the relevant orthogonal polynomials 
(see \S\ref{sec:gamma}).
 
\subsection{The Gaussian vs. the Gamma PDF} 
 
In the case of the {\bf Gaussian} (or normal $N(0,\sigma)$) PDF: 
\beq 
p(\del)~ = ~ p_G (\del)  
~ \equiv ~{1\over{\sqrt{2\pi \sigma^2}}}~\exp{\left[-{1\over{2}} 
\left({\del\over{\sigma}}\right)^2\right]}, 
\label{g1PDF} 
\eeq 
where $\sigma$ is the rms standard deviation,  
the only  parameter in this distribution. 
As the overdensity has to be positive $\rho>0$, we have that $\del <-1$, 
and a Gaussian PDF only makes physical sense  
when $\sigma \rightarrow 0$. For the Gaussian, $S_J=k_J=0$ for $J>2$. 
 
The {\bf Gamma} ---also called  negative binomial or Pearson Type III (PT3) 
PDF--- arises from  the Chi-Square distribution with $N$ 
degrees of freedom when $1/\sigma^2=N/2$ is taken to be a continuous 
parameter. This yields 
\beq 
p(\del)~ = ~ \phi(\del)  \equiv {(1+\del)^{\sigma^{-2}-1}  
\over \sigma^{2\sigma^2} \Gamma(\sigma^{-2})} ~\exp{\left(-{1+\del\over{ 
\sigma^2}}\right)}.  
\label{gamPDF}
\eeq 
The hierarchical coefficients in Eq(\ref{sj}) are constant for all 
values of the variance in this case and give $S_\J=(\J-1)!$.
These $S_J$ values are equal to those of a simple exponential distribution.  
The Gamma PDF (or similar ones) has been found to be useful at describing 
the galaxy distribution (see eg, Fry 1986, Elizalde \& Gazta\~naga 1992,
Gazta\~naga \& Yokohama 1993). 
   
\subsection{The Saddle Point Approximation} 
 
 The moment generating function summarizes all the information  
concerning the higher-order cumulants, as long as the series expansion in  
terms of the latter converges. In the majority of the cases this is true  
(one counterexample to this rule is the Lognormal distribution). Thus,  
one may reconstruct the PDF from the the moment generating function
in a consistent way as  
shown in Eq(\ref{PDF-mgf}). 

To obtain an asymptotic expansion of $p(\delta)$ for small $\delta$, we introduce  
the Legendre transform, 
\beq 
\bar{\delta} \equiv d\psi(t)/dt, \quad  G(\bar{\delta}) = \bar{\delta} t - 
\psi(t), 
\eeq 
where the convexity of $G(\bar{\delta})$ is related to that of $\psi(t)$.    
Replacing this in the original expression for $p(\delta)$, we are left with 
\beq 
p(\delta) =  \int_{G'=-i \infty}^{G'= +i \infty}
\hspace{-1mm} {{G'' d \bar{\delta}} \over {2 \pi}} 
\exp[{-\delta G'(\bar{\delta})\,+\,\bar{\delta}G'(\bar{\delta})\,-\,G(\bar{\delta})}] 
\eeq 
which is dominated by stationary points of the exponential at $\delta=\bar{\delta}$  
(for real finite $\delta$). The Saddle Point approximation of this integral is  
given in Fry (1985) and follows from the usual approach (Morse and  
Feshbach 1953):  
\beq 
p(\delta) \sim [G''(\delta)/{2\pi}]^{1/2}\, \exp[-G(\delta)]. 
\eeq  
Once normalized, the distribution reads as 
\beq 
p(\delta) = {{[G''(\delta)/{2\pi}]^{1/2}\, \exp[-G(\delta)]} \over  
{\int_{- \infty}^{ + \infty}{[G''(\delta)/{2\pi}]^{1/2}\, \exp[-G(\delta)]d\delta}}}. 
\eeq 
Provided one constructs the generating function $\psi(t)$ out of the irreducible 
moments, 
\beq 
\psi(t)= \sum_{n=2}^{\infty} {{\mu_n} \over {n!}} t^{n} 
 = {1 \over 2} \sigma^2 t^2 + {1 \over 6}S_3 \sigma^4 t^3 +   
{1 \over 24}S_4 \sigma^6 t^4 + \cdots 
\label{psit}
\eeq 
(where the latter equality shows the expansion in terms of the hierarchical 
amplitudes),  
there is a general development for $G(\delta)$ in powers of $\delta$ which is derived by 
inverting the $t$ variable in the Legendre transformation as to get $t=t(\delta)$, 
leading to  
\beq 
G(\delta) \approx \left[{1 \over 2}\delta^2\,-\,{S_3 \over 6} \delta^3\,+\,{1\over 8} 
\left({S_3}^2 \,-\,{S_4 \over 3} \right)\delta^4 +\,{\cal O} (\delta^5) 
 \right]\, {\sigma^{-2}}. 
\label{saddleg}
\eeq 
To get a proper expansion around the Gaussian PDF, we first need to arrange 
the exponential by factoring out the quadratic term. In the limit of 
small $\sigma$, we then get 
%\bea 
%\exp[-G(x)] &\approx& \exp \left\{{\sigma^{-2}} \left[{S_3 \over 6} x^3\,+ 
%\right. \right. \\ \nn &+& 
%\left. \left. \, {1\over 8} \left({S_4 \over 3}\,-\,{S_3}^2 
% \right)x^4 +\,{\cal O} (x^5) \right]\right\} 
%\cdot \exp \left[-x^2/(2 {\sigma^2})\right] 
%\eea 
%and, Taylor expanding the first exponential term before the Gaussian, we get 
\bea 
\exp [-G(\nu)] &\approx & \left[1\,+\,{1 \over 3!}S_3 {\nu}^3 \sigma \, 
+\,{1\over 4!}\left(S_4\,-\,3{S_3}^2 
\right){\nu}^4 \sigma^2  \right. 
\nn \\&+& \left. \,{10\over 6!}  {S_3}^2 {\nu}^6 \sigma^2 
\right] \, p_G (\nu) +\,{\cal O} 
\left( {\sigma^3}\right), 
\eea 
where $\nu \equiv  \delta/\sigma$.
Applying the same expansion to the denominator (the normalization) 
and recalling the property for Gaussian integrals 
\beq 
{1 \over \sqrt{2\pi}}\int_{- \infty}^{ + \infty}{{\nu}^n e^{-{{\nu}^2 \over 2}} 
d{\nu}} = \left\{ \begin{array}{ll} 
                        (n-1)!!  & \mbox{ if $n$ even}, \\ 
                           0  & \mbox{ otherwise}, 
                         \end{array}             
                        \right. 
\eeq 
we obtain a polynomial in 
powers of the small parameter ${\sigma^2}$ of the form: 
\bea 
&& \left[{\int_{- \infty}^{ + \infty}{d\delta [G''(\delta)/{2\pi}]^{1/2}\, \exp[-G(\delta)]}} 
\right]^{-1} \nn \\ && 
\simeq 1\,-\,\left\{ {150 \over 6!}{S_3}^2\,-\,{3 \over 4!}S_4 \right\} {\sigma^2} 
\,+\,{\cal O} \left( {\sigma^3}\right). 
\eea 
Finally, we multiply both developments derived above and keep terms up to  
${\sigma^3}$, to end up with 
\bea 
p(\nu) &\simeq& \left[1\,+\,{S_3 \over 3!}H_{3}(\nu)\sigma\,+\,\left\{ {1 \over 4!} 
S_4 H_4 (\nu) \right.\right. \\ \nn 
&+& \left.\left. \,{10 \over 6!} {S_3}^2 H_6 (\nu) \right\}{\sigma}^2 \right] 
\, p_G (\nu)
+\,{\cal O} \left({\sigma}^{3}\right), 
\label{edge} 
\eea 
$H_n (\nu)$ being the Hermite polynomials, 
\bea 
H_3(\nu) &=& \nu^3\,-\,3\nu, \nn \\\nonumber 
H_4(\nu) &=& \nu^4\,-\,6\nu^2\,+\,3, \nn \\ \nonumber 
H_5(\nu) &=& \nu^5\,-\,10\nu^3\,+\,15\nu, \nn \\ \nonumber 
H_6(\nu) &=& \nu^6\,-\,15\nu^4\,+\,45\nu^2\,-\,15, \\ \nonumber 
H_7(\nu) &=& \nu^7\,-\,21\nu^5\,+\,105\nu^3\,-\,105\nu, \\ \nonumber 
H_9(\nu) &=& \nu^9\,-\,36\nu^7\,+\,378\nu^5\,-\,1260\nu^3+945\nu, \ldots 
\eea 
which is the well-known (perturbative) Edgeworth series of a PDF up to third order.
Higher-orders in the Edgeworth series can be obtained by keeping higher-orders in 
the Taylor expansions of Eqs(\ref{psit}),(\ref{saddleg}). 
 
We stress that the latter expansion is derived under the assumption that the  
distribution is hierarchical, ie, the $S_J$ are independent of  
${\sigma}$. 
According to this, the Edgeworth expansion may be generalized to  
non-hierarchical PDFs whenever the scaling of $S_J({\sigma})$ is  
known and replaced in the generating function $\psi(t)$ from which the  
Saddle-Point approximation of $p(\delta)$ is built. 
 
Non-linear PT for Gaussian initial conditions 
predicts corrective ($\sigma$ dependent) terms to 
the leading order contribution to $S_J$
of the form $S_J = S_J^{(0)} + S_J^{(1)} \sigma^2 + {\cal O} (\sigma^4)$,
where $S_J^{(0)},  S_J^{(1)}$ are coefficients independent of $\sigma$. 
This must be taken into account to make consistent predictions from 
non-linear dynamics 
for the third-order term (or higher) in the Edgeworth series of the PDF
(see Bernardeau \& Kofman 1995).

The Edgeworth expansion ---or any other expansion
based on the symmetry around the peak of the PDF being approximated--- is only  
a good candidate for fitting the evolutionary picture of the density  
profile as a first order approach, because high-density (non-Gaussian) 
exponential tails develop in further stages of the non-linear evolution for 
arbitrary initial conditions. 
 
 The Edgeworth series has also been applied to non-linear transformations 
of the Gaussian process to fit the exponential tails observed in the  
simulations as the system evolves. The `skewed' Lognormal approximation 
put forward by Colombi (1994) is an example of this scheme which takes 
advantage of the apparently privileged role the Lognormal PDF plays among  
the non-Gaussian ones. This is suggested by the integration of the continuity  
equation in Lagrangian coordinates on one hand, and by the good fit to  
the observational PDF on intermediate scales (related to a $n=1$ spectral index 
for a power-law power spectrum), on the other hand. 
     
 Nevertheless, it is still lacking in the literature a well-defined expansion around 
a non-Gaussian PDF which  
may be better suited than the Gaussian to model the 
gravitational evolution of cosmic fluctuations in the weakly non-linear regime.
It is not clear yet whether the initial conditions of structure formation 
in the universe were Gaussian (as suggested by standard inflationary models), or not.
In the latter case, it is necessary to investigate expansions around non-Gaussian PDFs 
if one wants to describe clustering.
These issues have been our main motivation to introduce an alternative expansion 
around a non-Gaussian PDF.
 
In the next section we derive a general expansion 
around the Gamma distribution, ie., around an arbitrary exponential tail, 
making use of the completeness and orthogonality properties of the Laguerrre 
polynomials. They are formally analogous to the Hermite polynomials that 
appear in the Edgeworth series around a Gaussian.

\section{Expansion around the Gamma PDF} 
\label{sec:gamma} 
 
Our starting point here will be the expansion of the PDF in terms of the Gamma  
distribution, with a basis that will be given by generalized Laguerre  
polynomials (instead of the Hermite ones in Eq(\ref{edge})). That such 
an expansion makes sense becomes clear from the  
fact that the Gamma distribution is proportional to the measure associated with
this particular family of orthogonal polynomials. 
In this sense, the Gamma expansion is formally reminiscent of the Edgeworth
series as the main difference consists of replacing the Hermite polynomials 
by the Laguerre ones as the basis for the expansion. 
Notice however that while the Edgeworth series
might be built from the Gram-Charlier series (Cram\'er 1946), 
given by the Gaussian PDF and its derivatives (see Juszkiewicz et al. 1995),
here we cannot simply 
take successive derivatives of the Gamma distribution.
This is because, in doing so, we would 
actually fail to make contact 
with a consistent theory of orthogonal polynomials. 
In short, if the 
generalized Laguerre ones are to be used (and those are the only possibility
in the case considered), 
it turns out that the integration measure changes with the {\it order} of the 
generalized polynomial (and not just with the parameter of the family), 
and this would invalidate the whole approach. 
We should, by the way, recall the good properties of a well defined
 expansion in terms  of an orthogonal  basis of
polynomials, that are orthonormalized with respect to an scalar product 
defined by a fixed integration measure. This is a rigorously defined
mathematical theory, that in the case of the Gaussian measure happens
to coincide with the Taylor expansion in terms of the derivatives of the
function. In the case of the Gamma function, on the contrary, it turns out 
that the two
expansions do not coincide, and only the one in terms of orthogonal 
polynomials has rigorous mathematical sense. Thus, this must be the 
starting point in our approach, as the Taylor expansion (given in terms 
of the successive derivatives) must be relegated to a mere formal expansion 
lacking adequate convergence properties.

The key point in our scheme is to build a general and consistent
expansion in terms of the relevant
orthogonal polynomials (the Laguerre ones, in the present case), 
{\it not necessarily} 
given by derivatives of the parent PDF (ie, we do not make use of 
Equation (\ref{p1p2})). Bearing this in mind, we modify the approach 
based on the general Eq(\ref{p1p2}) and define an expansion of the PDF 
in terms of the Gamma distribution as follows: 
\beq 
p(\mu ) \equiv \sum_{n=0}^\infty c_n L_n^{(p-1)} (\mu ) \phi (\mu ), 
\label{21}
\eeq 
ie, we define an expansion for which all orthogonal polinomials
are well defined, with coefficients: 
\beq 
c_n= \frac{n!\Gamma (p)}{\Gamma (n+p)} \int_0^\infty p(\mu )  
 L_n^{(p-1)} (\mu )\, d\mu , 
\eeq 
being $\phi(\mu)$ the Gamma PDF: 
\bea 
\phi(\mu)d\mu &=& {1 \over {\Gamma(p)}}\mu^{p-1} e^{-\mu}d\mu, \\  
\mu &=& {x\,-\alpha \over \beta}\, \geq 0 . 
\eea 
This is actually a three-parameter ($p,\alpha$ and $\beta$) family of distributions  
out of which only one, $p$, is 
relevant for normalized variables (such as density fluctuations, $\del$). 
 
The generalized Laguerre polynomials we shall need are given by: 
\beq 
L_n^{(p-1)} (\mu ) = \sum_{k=0}^n \frac{(-1)^k}{k!} 
\displaystyle{\left(_{\ n-k}^{n+p-1} \right)} \mu^k, 
\eeq 
in particular, 
\bea 
L_1^{(p-1)} (\mu ) &=& p-\mu, \nn \\ 
L_2^{(p-1)} (\mu ) &=&  
\frac{p(p+1)}{2}-(p+1)\mu +\frac{\mu^2}{2}, \nn \\ 
L_3^{(p-1)} (\mu ) &=&  
\frac{p(p+1)(p+2)}{6}-\frac{(p+1)(p+2)}{2}\mu \nn \\ 
&+&\frac{p+2}{2}\mu^2  
-\frac{\mu^3}{6},   
\eea 
and the coefficients $c_n$ are easy to calculate: 
\bea 
c_0 &=& 1, \quad c_1 =c_2 =0, \nn \\ 
c_3 &=& 
 -\frac{\Gamma (p+1)}{\Gamma (p+3)} (S_3 -2!), \nn \\ 
c_4 &=& \frac{\Gamma (p+1)}{\Gamma (p+4)} \left[ (S_4-3!)-12(S_3 -2!)\right],  \\   
c_5 &=&  \frac{\Gamma (p+1)}{\Gamma (p+5)}  
\left[- (S_5-4!) + 20 (S_4 - 3!)-120(S_3 -2!)\right] . \nn 
\eea 
In general, one finds: 
\bea 
c_n&=& \frac{\Gamma (p+1)}{\Gamma (p+n)} \sum_{k=3}^n \left\{(-1)^k a_{n,k} 
 \left[ S_k -(k-1)!\right] \right. \nn \\
&& \left. + b_{n,k} 
 \left[ S_k -(k-1)!\right]^2+ \cdots \right\}. 
\eea 
It is clear that for $S_k=(k-1)!$ we recover the Gamma PDF, $p(\mu) = \phi (\mu )$. 
 
Note that for the first coefficients it holds 
\bea 
&& a_{n,n} =1, \nn \\  
&& a_{n,n-1} =n (n-1) , \nn \ldots 
%\\ 
%&& a_{n,n-2} =2! n (n-1) (n-2), \nn \ldots 
\eea 
We should stress  the fact that for higher orders (ie., for $c_6$ or higher) 
there appear linear terms in $p=1/\sigma^2$  which couple to those quadratic 
in $\left[ S_k -(k-1)!\right]$.
%, and that this process may go on (in this 
%the expansion is not different from the Edgeworth one). 
For instance,
\bea
c_6 &=& \frac{\Gamma (p+1)}{\Gamma (p+6)} 
\left[ (S_6 - 5!) - 30 (S_5-4!) + 300 (S_4 - 3!) 
 \right. \nn \\ 
&-& \left.1200 (S_3 -2!)
+ 10\, p \, (S_3 -2!)^2 \right] .
\eea 
Notice that an appropriate
expansion for small $\sigma$ will have contributions from these higher-order terms.  
In order to see clearly which is the parameter in the expansion (\ref{21}), and 
set up a comparison with the Edgeworth expansion, we express 
Eq(\ref{21})  in terms of  
the same variable, $\nu = \delta /\sigma$, so that
$\mu=1/\sigma^2 +\nu/\sigma$. Doing this, we have: 
\bea
p(\nu) &=& 
\left\{ 1 + \sum_{n=3}^\infty \frac{\Gamma (1+1/\sigma^2)}{ 
\Gamma (n+1/\sigma^2)} L_n^{(1/\sigma^2-1)} (\mu) \right. \nn \\ &&
\times \left. \sum_{k=3}^n (-1)^k a_{n,k} [S_k-(k-1)!] \right\} 
\phi (\nu) \nn \\ & =& \left\{   
1 + \sum_{n=3}^\infty \sigma^{n-2} \, F_n \, 
\right. \nn \\ && \times \left.
\sum_{k=3}^n (-1)^{n-k} a_{n,k} [S_k-(k-1)!]\right\} \phi (\nu), 
\label{32}
\eea 
with the $F_n=F_n(\nu, \sigma)$ being of the form: 
\beq
F_n  (\nu, \sigma) =
\frac{1}{n!} H_n(\nu) + {\cal Q}_n(\nu)\, \sigma + {\cal R}_n(\nu)\, \sigma^2 + 
{\cal O} (\sigma^3) 
\label{fn}
\eeq
with $H_n$ the Hermite polynomial of order n, and  
\bea 
&& {\cal Q}_3(\nu) = \frac{2}{3}-\nu^2 
\nn \\  &&
{\cal Q}_4(\nu) = \frac{\nu}{6} \left(7-3 \nu^2 \right)
\nn \\  &&
{\cal Q}_5(\nu) = \frac{1}{6} \left(-2 + 5 \nu^2 - \nu^4 \right)
\nn \\  &&
{\cal Q}_6(\nu) = \frac{\nu}{72} \left(-33 + 26 \nu^2 - 3 \nu^4 \right)
\nn \\  &&
{\cal R}_3(\nu) = \frac{\nu}{2} \left(5-\nu^2 \right), \ldots
\label{qr}
\eea
%\bea 
%&& F_3  (\nu, \sigma) = \frac{\nu}{6}(\nu^2-3) - 
%\left( \nu^2 -\frac{2}{3} \right)\, \sigma -
%frac{\nu}{2} (\nu^2-5) \, \sigma^2 
%+{\cal O} (\sigma^3), 
%\quad  
%\nn \\ && 
%F_4  (\nu, \sigma) = \frac{1}{24}(\nu^4-6\nu^2+3) -
%\frac{\nu}{6}(3\nu^2-7) \, \sigma
%+{\cal O} (\sigma^2), \nn \\  &&  
%F_5  (\nu, \sigma) = \frac{\nu}{120}(\nu^4-10\nu^2+15) -
%frac{1}{6} (\nu^4 - 5 \nu^2 +2)\, \sigma
%+{\cal O} (\sigma^2), \ldots
%\eea 
%Thus, in general, we obtain 
%\beq 
% F_n  (\nu, \sigma) = \frac{1}{n!} H_n(\nu) +{\cal O} (\sigma), 
% \eeq 
%with $H_n$ the Hermite polynomial of order n. 
To summarize, we see that what we have in Eq(\ref{32}) is  
in fact  an expansion in terms of 
the Gamma PDF in the perturbative limit (ie., when $\sigma$ is small), 
in formal analogy to the Edgeworth expansion (which uses the Gaussian as 
the parent PDF). 
The new expansion should presumably 
be much better suited than 
the Edgeworth expansion, to parametrize 
PDFs  with  exponential tails. 
 
\section{Comparison of the Edgeworth \&  Gamma expansions} 
\label{sec:compare} 
 
We can now compare the Gamma with the Edgeworth series.
We just have to
replace Eqs(\ref{fn}),(\ref{qr}) in Eq(\ref{32}), and express
the (third order) Gamma expansion as a series in $\sigma$,
the rms standard deviation. We can write it in the following
compact notation:
\bea 
{p(\nu)\over{\phi (\nu)}} &=& 1\,+\, H_{3}(\nu) \Delta_3\, ~\sigma \nn \\ 
&+& \left\{ H_{4} (\nu) ( \Delta_4 - 3 \Delta_3 )\, 
+ H_6(\nu) \Delta_3^2/2 + 6\,{\cal Q}_3(\nu) \Delta_3 \right\} {\sigma}^2  \nn \\ 
&+& \left\{ H_5 (\nu) (\Delta_5 -4 \Delta_4 + 6 \Delta_3) + 
H_7(\nu) (\Delta_4 \Delta_3 - 3 \Delta_3^2)  
\right. 
\nn \\ 
&+& 
\left.  
H_9(\nu) \Delta_3^3/6 + 24 {\cal Q}_4(\nu) (- 3 \Delta_3 + \Delta_4 ) 
\right. 
\nn \\
&+&  
\left. 
360  {\cal Q}_6(\nu) \Delta_3 + 6 {\cal R}_3(\nu) \Delta_3      
\right\} {\sigma}^3 + {\cal O} (\sigma^4)  
\label{gammaexp}
\eea 
where $\phi (\nu)$ is the Gamma distribution (see Eq(\ref{gamPDF})) for
the dimensionless variable $\nu$.
%\beq
%\phi (\nu) \equiv {s^{1/s} \rho^{s-1} \over \Gamma(s)} ~e^{-s\rho}
%\eeq
%with 
%$\rho \equiv 1+\nu \sigma$ and $s=1/\sigma^2$. 
The polynomials ${\cal Q}_n ,{\cal R}_n$ are given above, Eq(\ref{qr}), 
and the reduced 
moments of the $p(\nu)$ PDF, $S_J$, 
appear as differences with the moments, $S_J^{(p)}$ of
the parent PDF (over which we are expanding):
\beq
\Delta_J \equiv {S_J-S_J^{(p)}\over{J!}}.
\eeq
In particular, for the Gamma expansion, $S_J^{(p)}=(J-1)!$. 
Eq.[{gammaexp}] is the main result of this paper.

For the Edgeworth series around the Gaussian one has up to the same order,
%$S_J^{(p)}=0$,
\bea 
{p(\nu)\over{p_G (\nu)}} &=& 1\,+\, H_{3}(\nu) \Delta_3\, ~\sigma \nn \\ 
&+& \{ H_4 (\nu) \Delta_4 \, 
+ H_6(\nu) \Delta_3^2/2 \} {\sigma}^2  \\ \nn 
&+& \{ H_5 (\nu) \Delta_5 + 
H_7(\nu) \Delta_4 \Delta_3  
+ H_9(\nu) \Delta_3^3/6  \} {\sigma}^3, 
\label{edgexp}
\eea 
where $p_G (\nu)$ is the Gaussian PDF with $S_J^{(p)}=0$.

%Note how, at each order, interms of these $\Delta_J$
%the contribution from the higher
%order moment has the same coefficient and polynomial in both
%the Gamma and the Edgeworth expansions. But the lower order
%corrections are different and 
%we have to introduced some new polynomials:
%\bea
%G_4(\nu) &\equiv& -3\nu^4\,+\,12\nu^2\,-\,5, \\ \nonumber 
%G_5(\nu) &=& -4\nu^5\,+\,28\nu^3\,-\,32\nu, \nn \\ \nonumber 
%G_5'(\nu) &=& 6\nu^5\,-\,27\nu^3\,+\,21\nu, 
%\eea
%Thus the Edgeworth expansion involves higher order polinomials in
%$\nu$ (up order 9th to same order in $\sigma$).

By comparing the Gamma and the Edgeworth expansions given above, we see that
the Gamma expansion recovers all the terms
that appear in the Edgeworth expansion plus some corrective terms.
In general the Gamma expansion has, by construction, 
exponential tails and a better general behaviour than the
Edgeworth expansion, both on
the positivity of $p(\nu)$ and the variate itself, $\rho$.

Notice that,
as suggested by expansions Eqs(\ref{gammaexp}),(\ref{edgexp}),
the convergence of these series depends on the magnitude of the coefficients
$\Delta_J \, \sigma^{J-2}$  that weight the contribution from every
polinomial. This is also true for the Edgeworth expansion, where
the natural smallnes parameter is given by $S_J \, \sigma^{J-2}$ (see also
Juszkiewicz et al. (1995)). This can already be guessed from the Saddle
Point approach to the Edgeworth series. In particular (see
Eq(\ref{psit})), the cumulant generating function, $\psi(t)$, is expanded in terms
of powers of $\sigma t$, with coefficients given by the cumulants 
$k_J \equiv  S_J \, \sigma^{J-2}$.
Accordingly, in the Gamma expansion, the convergence depends on how
close the tail of the PDF is to an exponential one within the perturbative limit.       

%macro read gamma.sm 
\begin{figure*} 
\centering 
%\centerline{\epsfysize=10.truecm %\epsfxsize=15.truecm\epsfbox{plotpdfl.ps}} 
\centerline{\epsfxsize=9.truecm\epsfbox{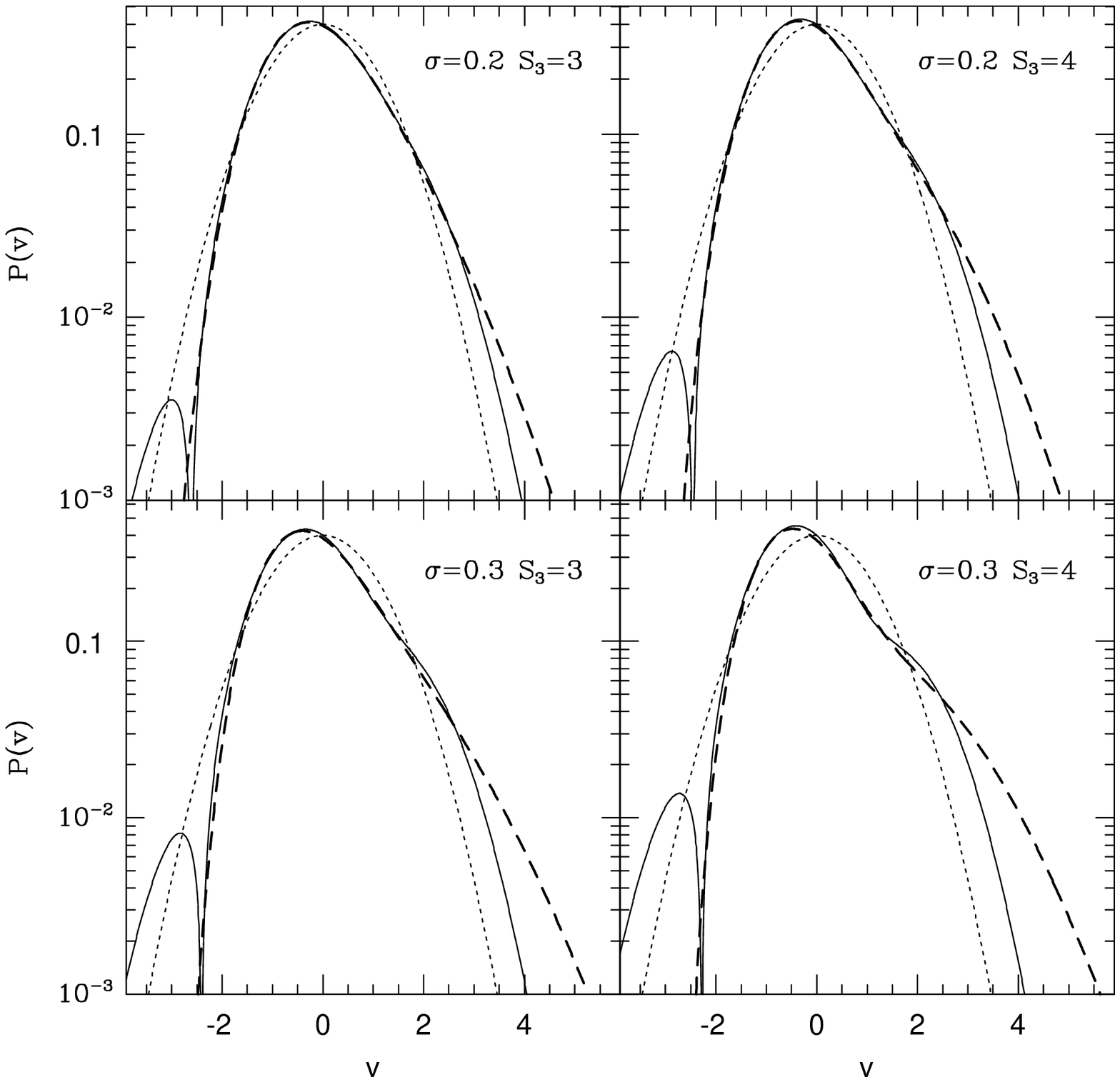}
\epsfxsize=9.truecm\epsfbox{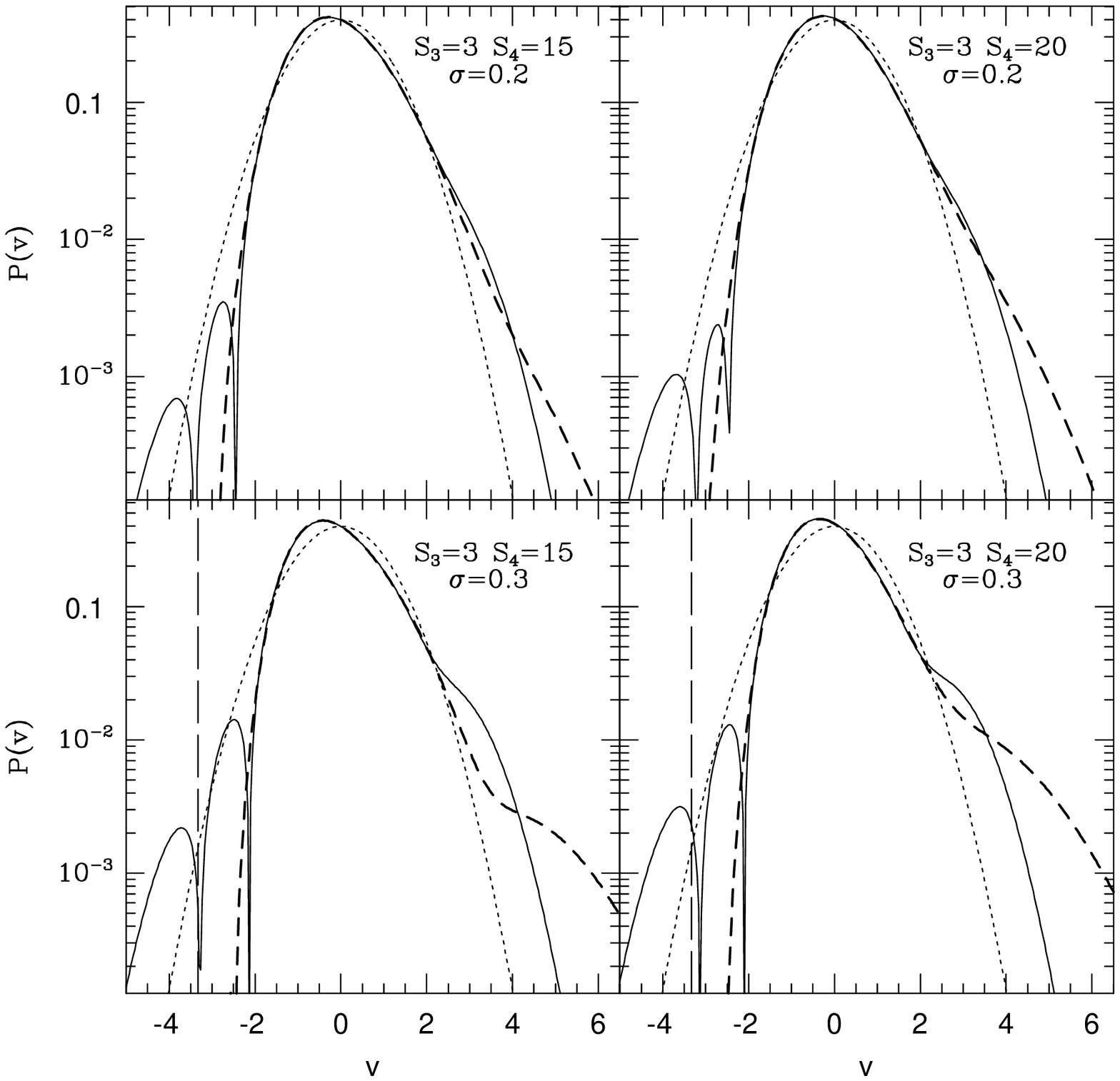}}
\figcaption[junk]{Comparison of the leading order (left panel) and second
order (right panel) Edgeworth  and  Gamma PDF expansions 
as  functions of $\nu \equiv \del/\sigma$ for several values of 
$\sigma$, and $S_J$ as labeled in the Figures. The  dotted, dashed and continuous   
lines correspond to the Gaussian PDF, Gamma and Edgeworth 
expansions, respectively.
\label{plotPDFl} }
\end{figure*}

Figure \ref{plotPDFl} (left panel) shows a comparison of the two expansions 
to leading order in $\sigma$,   
for $\sigma=0.2-0.3$ 
and $S_3=3-4$, as labeled in the Figures. 
For reference we also show the Gaussian distribution 
with the same $\sigma$ (dotted line). 
The Edgeworth expansion (continuous lines)  
quickly develops negative 
probabilities  (which are 
shown in absolute value) for negative values 
of $\nu \equiv \del/\sigma$.  
It is also clear 
how the Edgeworth has Gaussian tails dropping quickly 
to zero, while the Gamma expansion (shown as a dashed line) 
has exponential-type tails. 
 
Figure \ref{plotPDFl} also shows a comparison of the two expansions 
to the next order (second order) in $\sigma$,   
for $\sigma=0.2-0.3$, 
$S_3=4$, and $S_4=15-20$ as labeled in the Figures. 
The Edgeworth expansion (continuous lines)
shows now more negative values 
of the probability   
(shown in absolute value) for negative values of $\nu \equiv \del/\sigma$,
and large modulations for  $\nu>0$. For some range of parameters
and $\nu>0$ (ie. $ \del >0$),
the Gamma PDF also exhibits negative probabilities 
but they are typically 
much smaller than in the Edgeworth expansion.
This problem was less apparent 
in Fig. \ref{plotPDFl}, when working at the first order of the series
(left panel). 
 
The long-dashed vertical line corresponds to 
$\nu=-1/\sigma$, which marks the range of positive density: 
$\rho>0$ or $\nu > -1/\sigma$. 
This line is crossed both by the Gaussian 
PDF and the  Edgeworth PDF, indicating that these distributions 
cannot be used as physical distributions above some small 
value of $\sigma$.  
 
%An important observation is the fact that our procedure is universal, in the
%sense that 
%expansions around other distributions can, in principle, 
%be obtained in exactly the same way, by considering the generic expansion 
%Eq(\ref{p1p2}) and the orthogonal polynomials approach as
%described in  \S\ref{sec:expand} and \S\ref{sec:gamma}. 

%\begin{figure} 
%\centering 
%\centerline{\epsfysize=10.truecm %\epsfxsize=15.truecm  
%\epsfbox{plotpdf2l.ps}} 
%\caption[junk]{Comparison of the second order Edgeworth 
%and the Gamma PDF expansion 
%as functions of $\nu \equiv \del/\sigma$ for several values of 
%$\sigma$, $S_3$ and $S_4$ as labeled in the Figures. 
%The  dotted, dashed and continuous   
%line correspond to the Gaussian PDF, Gamma 
%and Edgeworth expansions, rspectively.} 
%\label{plotPDF2l} 
%\end{figure} 
 
\section{Comparison with simulations} 
\label{sec:nbody} 
 
As it is suggested by Eqs(\ref{gammaexp}),(\ref{edgexp}), differences in 
both expansions might be slight, specially around the peak of the PDF,
as to first order both expansions are formally equivalent. 
Thus, which one best fits numerical results
is a matter of careful analysis.  
The Edgeworth 
expansion has been shown to provide a very good approximation 
to model the PDF resulting from the weakly non-linear gravitational growth from small 
Gaussian initial fluctuations. We next compare both expansions with 
N-body simulations to see if we find significant deviations in their predictions. 
 
We measure the PDF in 10 realizations of  
SCDM simulations, $\Omega=1$ and $\Gamma=0.5$,
with $L=180 \Mpc$ and $N=64^3$  
particles and $\sigma_8=1$ (Croft \& Efstathiou 1994).
We consider several smoothing radius  
which correspond to different values of the variance, 
$\sigma^2$. The errors 
correspond the rms standard deviation in the 10 realizations. 
 
Figure \ref{PDFga09} corresponds to $\sigma \simeq 0.2$. As can be 
seen in the plot both expansions including the second order 
produce very similar results,
within the error bars. Note also that they both are significantly  
different from the Gaussian result (dotted line).  
Similar results are obtained for lower values of $\sigma$.

%\begin{figure}
%\centering 
\centerline{\epsfysize=8.5truecm %\epsfxsize=15.truecm  
\epsfbox{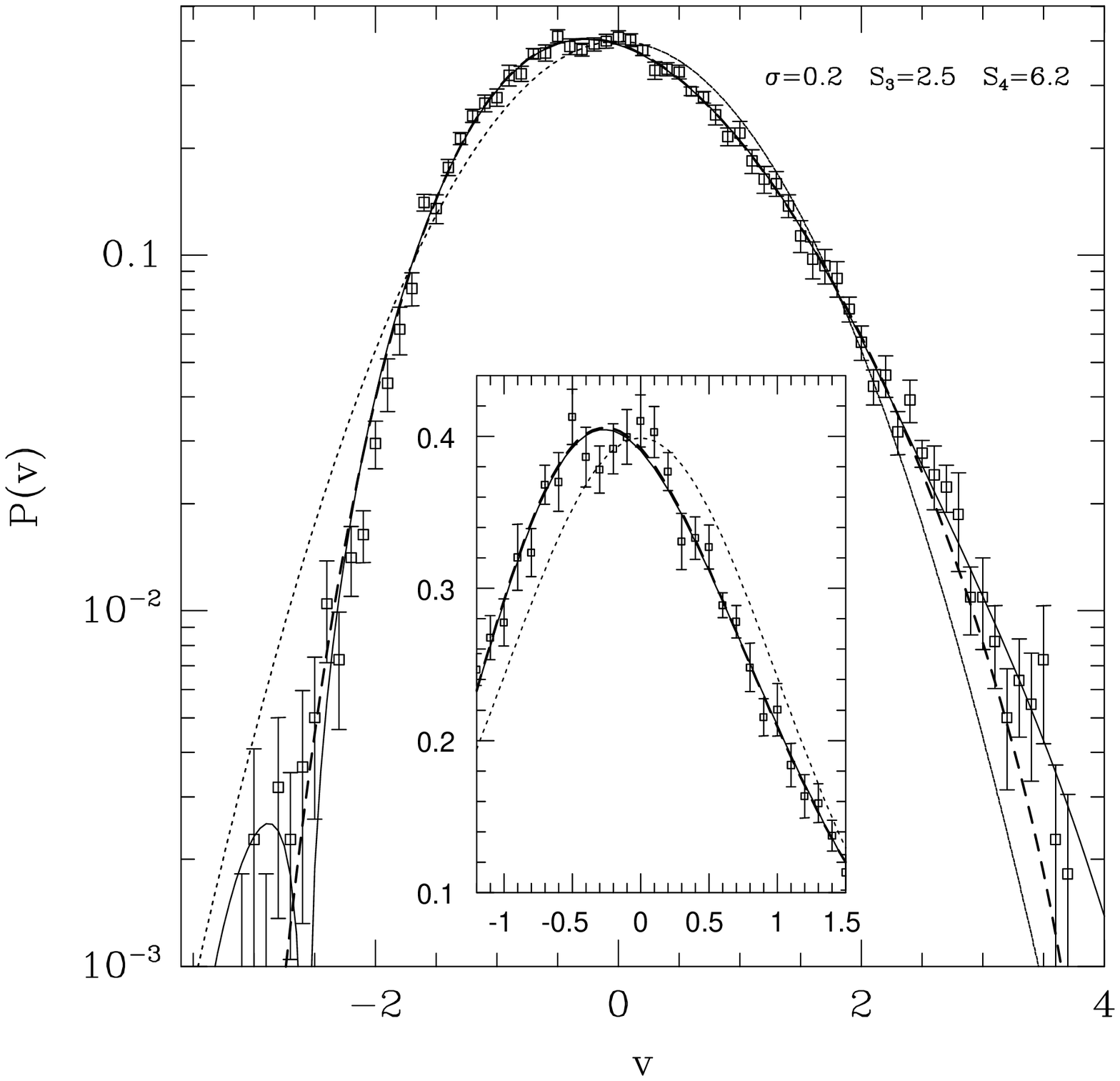}} 
\figcaption[junk]{Comparison of gravitational simulations with 
the second order Edgeworth and Gamma PDF expansions 
as a function of $\nu \equiv \del/\sigma$. We use as parameters the 
measured values of 
$\sigma$, $S_3$ and $S_4$ as labeled in the Figure.  
The  dotted, dashed and continuous   
lines correspond to the Gaussian, Gamma and Edgeworth distribution
expansions. The inset shows a detail around the peak in linear 
scale.
\label{PDFga09} }
%\end{figure} 
 
%\begin{figure} 
%\centering 
%\centerline{\epsfysize=8.5truecm %\epsfxsize=15.truecm  
%\epsfbox{pdfga07.ps}} 
%\caption[junk]{Same as Fig. \ref{PDFga09} for $\sigma \simeq 0.5$
%and including terms in the expansion up to third order in $\sigma$.}
%\label{PDFga07} 
%\end{figure} 

%Figure \ref{PDFga07} corresponds to $\sigma \simeq 0.5$. As can be 
%seen in the plot both expansions including third order 
%produce very similar results, 
%within the errorbars, around the peak. For $\nu \ga 3$ both
%expansions seem to break (at this order), with the Gamma
%showing large negative values. This is not surprissing as
%$\sigma \simeq 0.5$ is already quite large. The values of
%the PDF on the tail, for this relatively large value of
%$\sigma$, depend strongly on the values of $S_J$ used
%in the expansion..
%Also note the negative values of the Edgeworth for $\nu<-2$. 

\begin{figure*} 
\centering 
\centerline{\epsfysize=8.5truecm %\epsfxsize=15.truecm  
\epsfbox{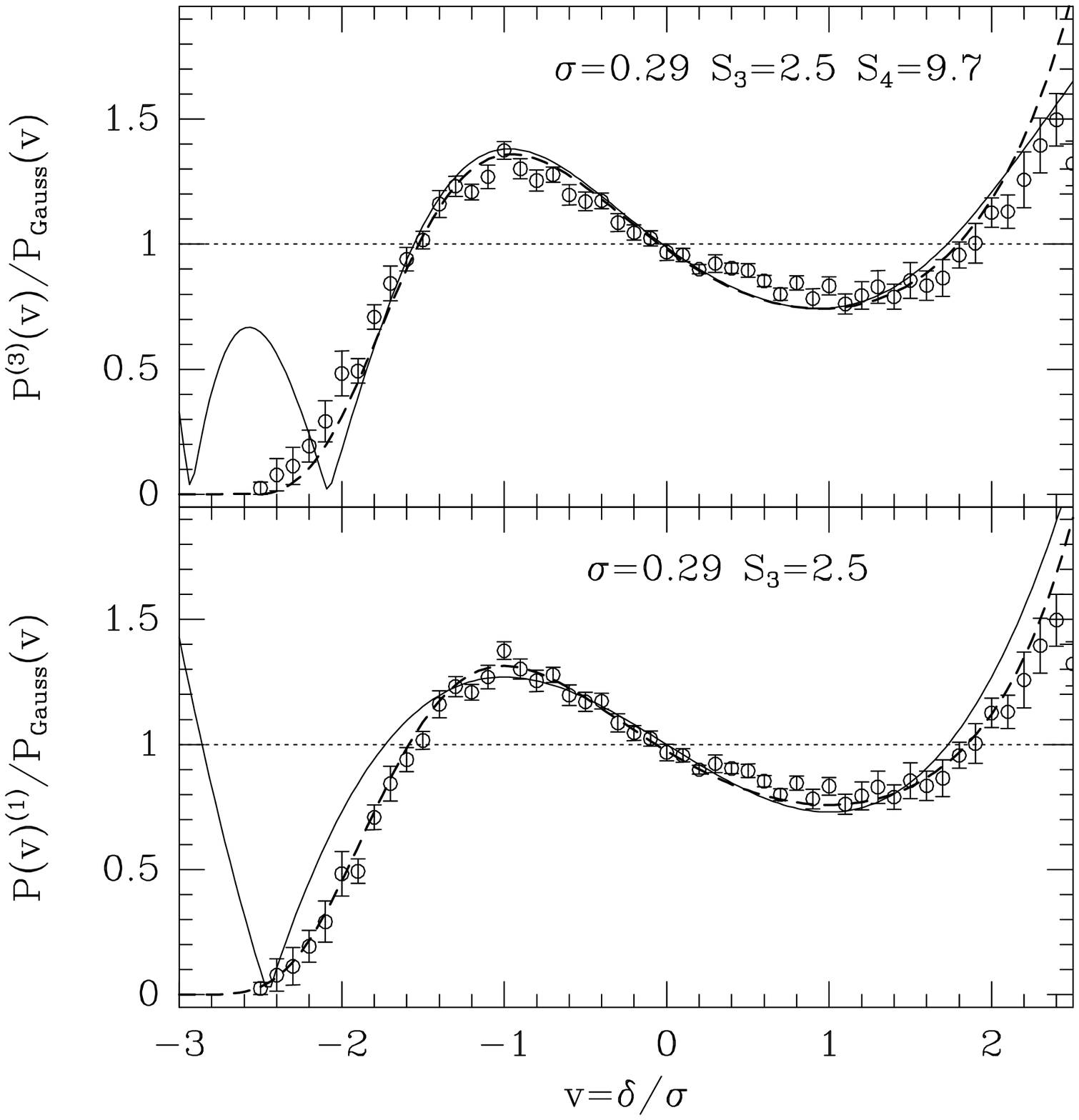}\epsfysize=8.5truecm %\epsfxsize=15.truecm  
\epsfbox{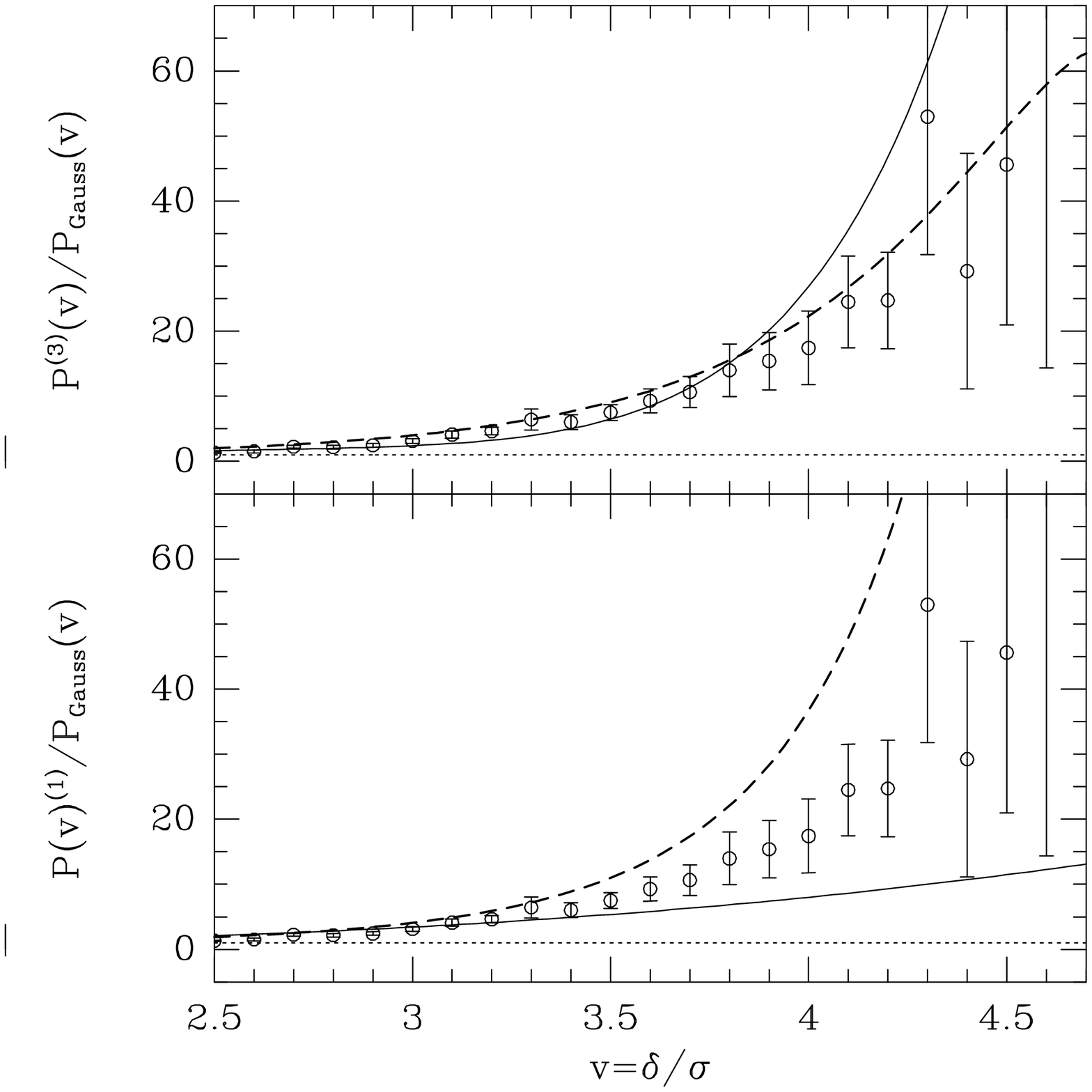}} 
\figcaption[junk]{Deviations from the Gaussian PDF for both expansions and in
N-body simulations (symbols). 
The lower panels displays results for the first order. 
The upper panels 
show the expansions including the third-order terms.
The solid line is given by the Edgeworth series while the dashed one 
shows the Gamma expansion. The left and right panels show different
ranges in $\nu$.
\label{PDFra08}}
\end{figure*} 

When $\sigma$ is small, $\sigma \la 0.3$, using the measured N-body 
(non-linear) values of $\sigma$ and 
$S_J$ as parameters for the PDF expansion, it 
yields very similar numbers, within the error bars, to the ones 
obtained using the corresponding 
linear $\sigma$ and non-linear PT predictions for $S_J$.
The latter is in agreement with what was suggested by  
Juszkiewicz et al. (1995), Bernardeau \& Koffman (1995). 
For larger values of the {\it rms} fluctuation, $\sigma \ga 1$, differences in the PDF 
when using perturbative or non-linear parameters become
larger.

%\begin{figure} 
%\centering 
%\centerline{\epsfysize=9.truecm %\epsfxsize=15.truecm  
%\epsfbox{pdfra08b.ps}} 
%\caption[junk]{Same as Figure \ref{PDFra08} 
%for large values of $\nu$.}
%\label{PDFra08b} 
%\end{figure} 

In Figure \ref{PDFra08}  we can see that 
for $\sigma \simeq 0.3$
the Gamma expansion
seems to be in better agreement with the numerical results than the Edgeworth
specially for negative values of $\nu$.
It is seen that
both expansions yield similar results, except for the negative tail which
is better recovered by the Gamma expansion again. 
The Gamma distribution seems to perform 
slightly better also around $\nu \simeq 1-5$ as can be
seen in Figure \ref{PDFra08} (right panel).
For small $\nu$ the Edgeworth series
yields relatively large negative probabilities of the order
$P(\nu) \simeq 0.01$.
Thus, in this case it would be better to use the  
Gamma expansion if our priority is a better overall behaviour. This 
needs not  be rigorously generic and the situation could change if 
we explore a different domain of the parameter space. It mainly
depends on how large the values of $\Delta_J$ are for the case
of interest. 
Also note that larger values of $\sigma$ require higher  
orders in the expansions, which could then change their
relative performances. 
   
Figure \ref{PDFc20} illustrates what happens for larger values of
 $\sigma$, where these expansions are expected to break down. 
Only the first order is shown, but
the agreement does not improve for higher orders. 
In this regime, $\sigma \simeq 1$,
the expansions do not work well, as expected, and one has to
use some different approach to model the PDF.
The short-dashed line shows the results of one such model:
the non-linear Spherical Collapse model 
(from Fig.7 in  Gaztanaga \& Croft 1999). Errobars in the
simulations are comparable to the size of the symbols.
The Spherical Collapse model for the 
evolution of density perturbations can be used as 
a local Lagrangian mapping to relate the initial
and evolved fluctuation. The only input required is the linear variance
$\sigma$ and its slope $\gamma$ (eg see Fosalba \& Gazta\~{n}aga 1998a,b,
Gazta\~{n}aga \& Fosalba 1998).
Notice however that, while the Edgeworth and Gamma
expansions could be used to
model an arbitrary PDF, given its moments $S_J$, the Spherical Collapse
is only intended to model
gravitational dynamics of cosmic fluctuations.

%\begin{figure} 
%\centering 
\centerline{\epsfysize=8.5truecm %\epsfxsize=15.truecm  
\epsfbox{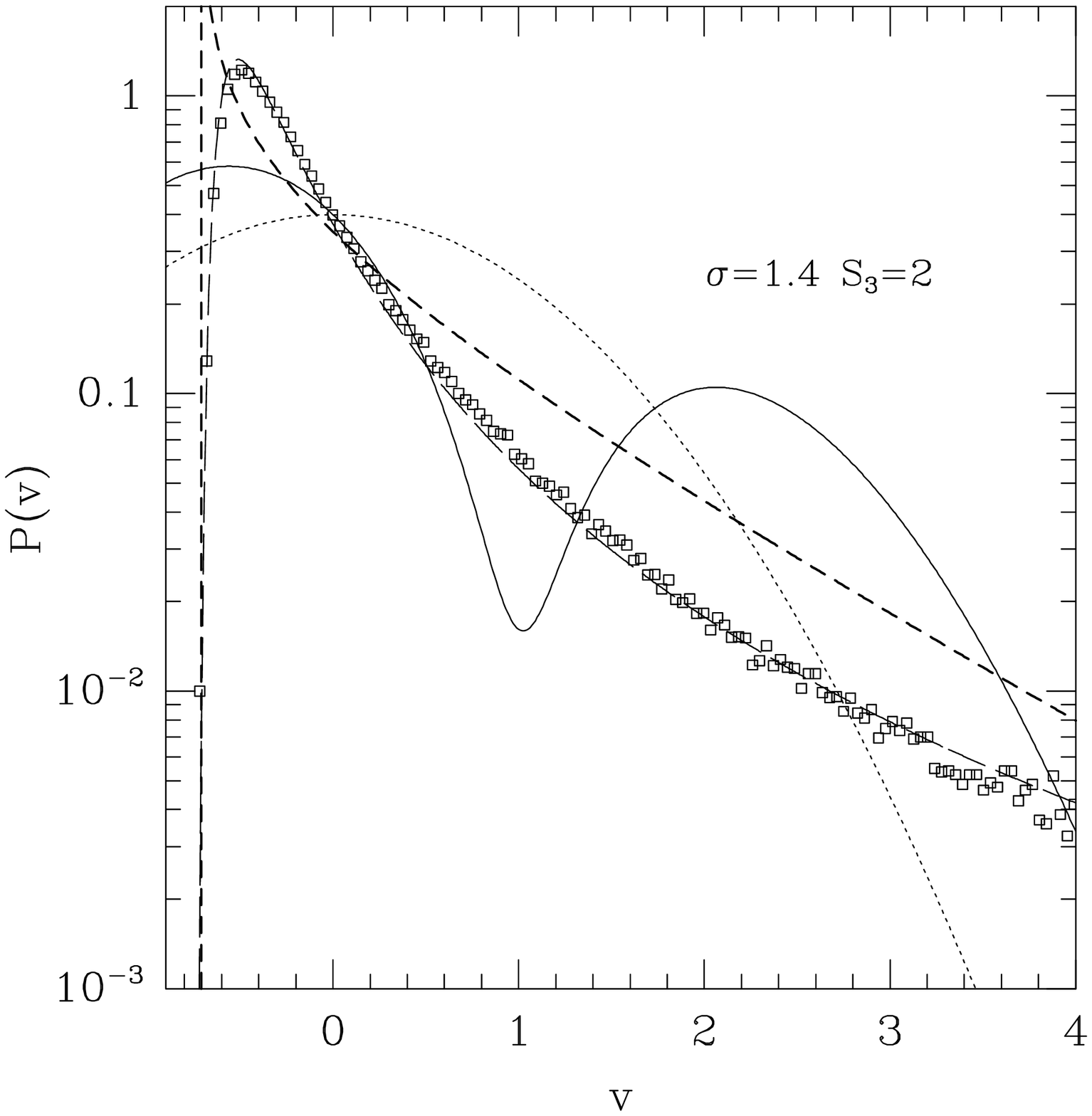}}
\figcaption[junk]{Comparison of Nbody results (symbols) with the
Gaussian (dotted line), the Edgeworth series
(continuous) and the Gamma expansion 
(short dashed line). Both expansions are given to first order. 
The long-dashed line following closely the simulation points corresponds
to the spherical collapse model.
\label{PDFc20}}
%\end{figure} 

%\newpage

\section{Discussion \& Conclusion}
\label{sec:conc} 

In this paper we have introduced 
the Gamma expansion, as an alternative 
to the well-known Edgeworth series, to model the gravitational evolution of the
density field on large-scales in terms of its lower order moments
$S_J \equiv {k_J /{k_2^{J-1}}}$ (see Eq(\ref{sj})).
The search for an alternative expansion is motivated by the fact that
the Gaussian PDF, which is used as the parent distribution for the Edgeworth series,
is not strictly well-defined for describing positive variates, such as the density
field. As a consequence, the expansion built out of the Gaussian exhibits
undesirable properties, namely, it predicts negative probabilities and 
allows for negative densities.    
In \S\ref{sec:expand} we have provided 
an independent derivation of the Edgeworth series based
on the saddle point approximation to the Legendre transform of the PDF.

In the case of the Gamma expansion (see \S\ref{sec:gamma}), 
the basis for the series is given by the generalised
Laguerre polinomials. 
These are the counterparts to the Hermite  polinomials which appear for 
expansions around the Gaussian, and allow for the construction of a consistent 
series in the perturbative limit (when the variance is small).
The coefficients in both expansions can be writen in terms of 
$\Delta_J \equiv {(S_J-S_J^{(p)}) /{J!}}$, where the reduced 
cumulants of the PDF, $S_J$, 
appear as differences with the corresponding cumulans, $S_J^{(p)}$ of
the parent PDF (over which we are expanding), ie. $S_J^{(p)}=0$
for a Gaussian PDF, or $S_J^{(p)}=(J-1)!$ for the Gamma PDF (see 
Eq(\ref{gammaexp}) and Eq(\ref{edgexp})).
The Gamma expansion  recovers all the contributions that appeared
in the Edgeworth series, as functions of $\Delta_J$,
plus some corrective terms, for second order in $\sigma$
or higher. 
The convergence of the series is safe as long as the
PDF develops a tail close enough to exponential, ie. when 
$\Delta_J$ are small.
Gravitational clustering predicts the appearence of such exponential tails in
the weakly non-linear evolution of the cosmic density field, at least for
Gaussian initial conditions. 
Therefore, the expansion introduced in 
this paper should constitute a good candidate to 
properly model clustering on large scales. 

We have carried out a detailed comparison of the performance of the Edgeworth
and the Gamma expansions in the perturbative
 regime with respect to cosmological N-body simulations with
$\sigma \la 0.4$. We have found that
they both  yield a very similar agreement with numerical results 
around the peak of the distribution. 
The Gamma expansion provides a better general match to the PDF on the tails.
In particular, the negative density tail measured in 
N-body simulations is accurately recovered from the Gamma expansion, 
unlike the case of the Edgeworth series.
Nevertheless, in general, the performance of the expansions
depend strongly on the values of $\Delta_J$. 
The agreement is better for the expansion which has its
parent moments $S_J^{(p)}$ closer to those of the PDF we want to model.
In other words,  
the smaller the $\Delta_J$, the better the behaviour of the expansion of the PDF.

In summary, the Gamma expansion provides an interesting
alternative to the Edgeworth series without introducing  any additional
mathematical entanglement. 
Both expansions have the same inputs (the
cumulants of the PDF we want to model) and outputs (the recovery of the full PDF)
and similar expressions (see Eq(\ref{gammaexp}) and Eq(\ref{edgexp})).
The proposed Gamma expansion is better suited for describing a real PDF,
because it always yields positive  
densities and the PDF is effectively positive-definite. 
It is also possible to expent the Gamma expansion to a  multivariate form,
using a generalization of the Laguerre polinomials.
Finally, we stress that many of the arguments presented here are 
rather generic and they might be useful when adressing the problem 
of modeling PDFs in other contexts.

%\section* 
\medskip 
\acknowledgements
This work has been 
supported by CSIC, DGICYT (Spain), project 
PB96-0925, and CIRIT (Generalitat de Catalunya), grant 1995SGR-0602.

\section{References} 
%\begin{references} 
 
\def\refe {\par \hangindent=.7cm \hangafter=1 \noindent} 
\def\aj {ApJ, } 
\def\na {Nature, } 
\def\aa {A\&A, } 
\def\prl {Phys.Rev.Lett., } 
\def\prd {Phys.Rev. D, } 
\def\phr {Phys.Rep. } 
\def\ajs{ApJS, } 
\def\mn {MNRAS, } 
\def\apl {ApJ Lett.), } 
 
\refe Amendola, L., 1996, \mn 283, 983 
\refe Balian, R., Schaeffer, R., 1989, \aa 220, 1
\refe Bernardeau, F., 1992, \aj 392, 1 
\refe Bernardeau, F., Kofman, L., 1995, \aj 443, 479
\refe Blinnikov, S., Moessnew, R., 1998, \aa 130, 193
\refe Coles, P., Jones, B., 1991, \mn  248, 1
\refe Colombi, S., 1994, \aj 435, 536
\refe Croft, R.A.C., \& Efstathiou, G., 1994, \mn 267, 390  
\refe Elizalde, E., Gazta\~{n}aga, E., 1992, \mn 254, 247 
\refe Fosalba, P., \& Gazta\~naga, E., 1998a, \mnras, 301, 503
\refe Fosalba, P., \& Gazta\~naga, E., 1998b, \mnras, 301, 535
\refe Fry, J.N., 1985, \aj 289, 10 
\refe Fry, J.N., 1986, \aj 306, 358
\refe Gazta\~{n}aga, E., Croft, R.A.C., 1999, \mn in press
\refe Gazta\~naga, E., \& Fosalba, P., 1998, \mnras, 301, 524
\refe Gazta\~{n}aga, E., Yokoyama, J., 1993, \aj 403, 450
\refe Juszkiewicz, R., Weinberg, D.H., Amsterdamski, P., 
Chodorowski, M., Bouchet, F.R., 1995, \aj 442, 39  
\refe Kim, R.S., Strauss, M.A., \aj 493, 39
\refe Kendall, M., Stuart, A. Ort, J.K., 1987,  
Kendall's Advanced Theory of Statistics, 
Oxford University Press %, New York) 
\refe Kofman, L., Bertschinger, E., Gelb, J.M., Nusser, A., \& Dekel, A., 
1994, \aj 420, 44 
\refe Saslaw, W.C., Hamilton, A.J.S., 1984, \aj 350, 492 
\refe Pen, U., 1998, \aj 504, 601 
\refe Popa, L., 1998, New Astronomy 3, 563
%\refe Peebles, P.J.E., 1980, {\it The Large Scale Structure of the  
%Universe:} Princeton University Press 

%\end{references} 

\end{document}